# On the competition between the phenomena involved in the aerosol dynamics

A. Michau, K. Hassouni, C. Arnas, G. Lombardi.

*Abstract*—
We made use of a combination of three models that describe plasma equilibrium, cluster kinetics and aerosol dynamics in order to investigate solid carbonaceous dust formation in a DC gas discharge where the carbon source is provided by a graphite cathode sputtering. This enabled us to determine the time-evolution of some of the particle characteristics, e., g. total density, mean-diameter, average-charge as function of the sputtering yield.  We showed that for few minutes discharge duration, the particle density reaches $10^7$-$10^8$ cm$^{-3}$, the particle mean diameter is approximately 40 nm and the particle size distributions are bimodal for all the investigated conditions. We also showed that the variation of the sputtering yield affects in a quite unexpected way the interplay between the phenomena involved in the aerosol dynamics. Basically, larger sputtering yields result in larger density and coagulation rate on one hand, and in a limited nucleation and surface sticking kinetics on the other hand.

## I. INTRODUCTION

Although solid particle formation is routinely observed in low pressure DC discharges [1] [2], very little effort has been devoted to the understanding of the processes that lead to their formation. Recently we showed, in the case of graphite cathode argon DC discharge, that particle formation is induced by molecular growth and a subsequent electron attachment of the sputtered carbon species [3].It results in the formation of negatively charged clusters that are trapped in the field reversal of the negative glow and can therefore undergo further growth that end up with solid particle nucleation. This growth mechanism was assessed through a first model that did not take into account the dusty plasma effect, *i.e.* the coupling between the discharge equilibrium and the particle aerosol dynamic was not taken into account. Nevertheless, this first model enabled us showing that field reversal effect in the NG plays a key role as far as particle nucleation in DC sputtering discharge is concerned [3]. Then, in a second step, we used a self consistent dusty plasma model that couples the discharge equilibrium, the molecular growth kinetic and the particle aerosol dynamic described through the number-density, the mean-diameter and the average-charge of the particles. We showed that field reversal effect still enables particle formation when the coupling between the space charge field and particle formation is taken into account [4]. Then, in a third step we investigated in more details the aerosol dynamic by solving the full master equations for the particle size distribution. We showed in particular that the size distribution is 'bimodal' in nature with two strong maxima located in the vicinity of the nucleation zone, i.e., 1 nm-diameter particle, and around the average diameter [5].

The space-time distribution of the particle population is governed by the interplay between the nucleation, the growth through cluster deposition on the particle surface and the coagulation [5]. The kinetics of these phenomena depends on some discharge and plasma-surface interaction parameters such as the sputtering yield and the sticking coefficient of the carbon clusters on the particle surface. In this article we examine how these parameters affect the interplay between the phenomena involved in the aerosol dynamic and the resulting particle density and size-distribution in the dusty plasma. In order to simplify the discussion, we do not consider in this study the dusty plasma effect, i.e., coupling between carbon charged species and discharge equilibrium. We limit ourselves to discharge durations where this coupling can be neglected.

The paper is organized in five sections, including this introduction. In section 2, the investigated discharge is briefly presented. The main features of the models used to describe the investigated dusty plasmas are described in section 3. In section 4, we show the results obtained especially on the time-space evolution of the particle cloud characteristics, *i.e.* density, average diameter, charge, size-distribution, and on the effect of the sputtering yield on these parameters. We interpret the obtained results in terms of the interplay between the phenomena involved in the aerosol dynamic. The main conclusions that may be drawn from this work are given in section 5.

## II. INVESTIGATED DISCHARGE

We consider an argon DC discharge generated between two parallel electrodes with a 14 cm gap. The cathode is made of graphite and the grounded anode in stainless steel. The pressure is 60 Pa and the discharge current is set to 80 mA which corresponds in these conditions to a -550 V voltage. The cathode is sputtered by argon ions and fast argon atoms

A. Michau, K. Hassouni, and G. Lombardi work at LSPM-CNRS, Université Paris 13, Sorbonne Paris Cité, 99 avenue Jean-Baptiste Clément 93430 Villetanesue, France (e-mail: armelle.michau@lspm.cnrs.fr; hassouni@lspm.cnrs.fr; lombardi@lspm.cnrs.fr).
C. Arnas works at LPIIM Aix-Marseille université, CNRS, Campus Saint Jérôme, 13397 Marseille, France (e-mail: cecile.arnas@univ-amu.fr).

produced in the cathode sheath [6]. For these conditions, where the energy of theses argon species are around 70 eV, carbon atoms and, to a lesser extent, small $C_2$ and $C_3$ clusters are produced with a sputtering yield of ~1%.

An experimental study was performed on the particles produced by this discharge [7]. After 300 s of discharge duration, the particles collected showed an average diameter of approximately 40 nm and a density as high as $10^8$ cm$^{-3}$ [8].

In the next section, we briefly present the model used for the description of the cluster and aerosol dynamics in a DC discharge.

### III. Particle growth scenario and modeling principle

Particle formation starts with the sputtering of the graphite cathode that releases C and, to a lesser extent, $C_2$ and $C_3$ neutral clusters. Clusters are transported in the discharge. These molecular species can grow by collision with other molecular species. This molecular growth phase is governed by aggregation processes (1-3), charge transfer (4) and attachment (5). Positively charged cluster can be produced but they are rapidly accelerated back in the cathode fall and lost at the cathode surface. Their residence time in the discharge is very low and they are therefore not involved in the growth process.

Fig 1 shows the cluster growth pathway starting with the sputtering of small neutral clusters and ending by solid particle nucleation. For the low pressure values encountered in the discharge considered, the neutral clusters diffuse very rapidly to the wall where they are lost by surface reaction. The residence times of these neutral species are therefore not high enough to enable clusters growth by aggregation and subsequent nucleation. Particle production is therefore only possible through negative clusters that can be trapped in the positive plasma potential. For these negative clusters, aggregation from C until $C_{30}$ can takes place significantly and particle nucleation can occurs near the field reversal position [3].

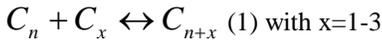

$$C_n + C_x \leftrightarrow C_{n+x} \text{ (1) with x=1-3}$$

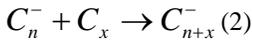

$$C_n^- + C_x \rightarrow C_{n+x}^- \text{ (2)}$$

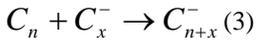

$$C_n + C_x^- \rightarrow C_{n+x}^- \text{ (3)}$$

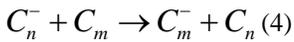

$$C_n^- + C_m \rightarrow C_m^- + C_n \text{ (4)}$$

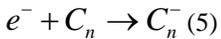

$$e^- + C_n \rightarrow C_n^- \text{ (5)}$$

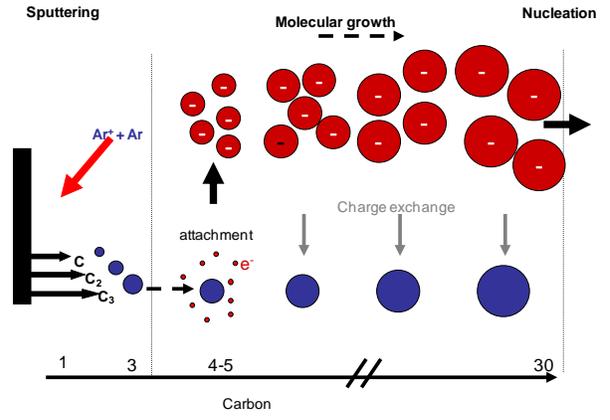

**Fig. 1 : schematic of the cluster growth pathway**

Therefore, for a realistic molecular cluster model, neutral and negative clusters are taken into account. These clusters undergo a molecular growth from the smallest sputtered species to the largest considered, i.e. the Largest Molecular Edifice, which is a model parameter. Once the molecular species reach this upper size limit we assume that they instantaneously lead to a solid particle nucleation, i.e. nucleation. Using this model, the time evolution of the cluster densities is governed by the following continuity equations :

$$\frac{\partial n_i^z}{\partial t} = -\nabla(-D_i \vec{\nabla} n_i^z + \mu_i.z.\vec{E}) + W_i^z$$

Where $n_i^z$, $D_i$, $\mu_i$ and $W_i^z$ represent the density, diffusion coefficient, mobility and net production rate of the $i^{th}$ cluster of charge $z$. $z$ may be either 0 or -1. E is the electric field.

Although the model can solve for the coupling between discharge equilibrium and charged cluster and particle formations [4], we limit the resolution to discharge durations where charged clusters and particles do not affect the discharge equilibrium. This allows for a straightforward analysis of the interplay between the different particle growth mechanisms. The discharge model we used consider three electron groups : (1) the very fast electrons that are produced and accelerated in the sheath,(2) the electron population that results from the relaxation of the fast electrons down to the first excitation threshold of argon in the glow, and (3) the very low energy electrons that are produced in the glow. Basically the first electron group is responsible for ionization, the second electron group ensures current continuity inside the Faraday dark space and the last group of electrons which is by far the major one determines the plasma density in the negative flow and the Faraday dark space.

It is necessary to determine the characteristics of each region of the DC discharge in terms of electron density and electron distribution function. The first module of the discharge model describes the sheath structure and is essentially inspired from the non-local approach proposed by Kolobov and Tsendin and based on the solution of the fast electron momentum [9]

Once the sheath dimension and the electric field variation in the sheath are determined we perform a Monte Carlo simulation to determine the evolution of the relative ionization rate in the sheath and the negative glow [3].

Once the absolute ionization rate is determined in the sheath and in the negative glow, the (cold) electron density in the negative glow and the Faraday dark space is determined by solving the 1-D transport equation for the cold electrons as described in [3]

The electric field in the NG is determined assuming a Boltzmann equilibrium with the cold electrons.

Once the plasma characteristics are estimated from the plasma model, we solve for the cluster continuity equations coupled to three balance equations that govern the particle number density, mass density and average charge [3]. This allow us to obtain the cluster densities and the nucleation rate taking into account the cluster losses at the particle surface.

Then in a second step the space-time distributions of these parameters are used in a detailed aerosol dynamics model with the objective to determine the size distribution of the solid particles. We used the so-called sectional model [9, 10] which allows for a good description of particle size distribution and particle coagulation growth process. In such a method, the size distribution is divided into several bins or sections as presented in Fig. 2.

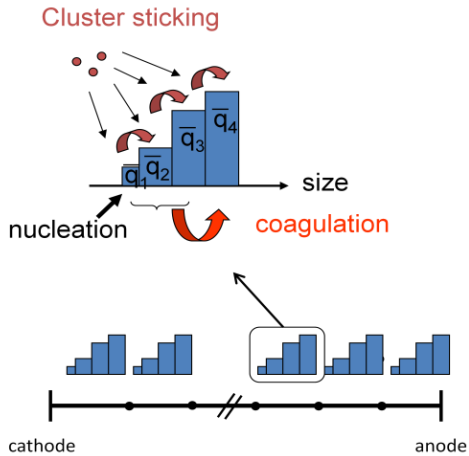

**Fig. 2 : principle of sectional model**

For each section we solve the following balance equation of volume conservation taking into account coagulation, cluster sticking at the particle surface, nucleation and transport.

$$\frac{\partial Q_l(t)}{\partial t} = \frac{\partial Q_l(t)}{\partial t}\bigg|_{coag} + \frac{\partial Q_l(t)}{\partial t}\bigg|_{sticking} + S_{nucleation} - \nabla F_l$$

where $Q_l$ is the particle volume in section l

The particle charge distribution is considered to follow a Gaussian distribution [11]:

$$\psi(q, q_p) = \frac{1}{\sigma\sqrt{2\pi}} \exp\left[-\frac{(q - q_p)^2}{2\sigma^2}\right]$$

$$\sigma = f\left(\frac{T_e}{T}, \frac{q_p}{d_p}, d_p\right)$$

Consequently, we only solve for the averaged charge and then use the previous Gaussian distribution to determine the instantaneous particle charge distribution for each space-time position and each section using. For each section (or particle size), the average charge is determined from a balance equation that takes into account the charging dynamics through electron and ion collections using OML Theory), transport, nucleation, coagulation and sticking of clusters. This may be written:

$$\frac{\partial \rho_l}{\partial t} = S^q_{e,i(OML)} + S^q_{transport} + S^q_{nucleation} + S^q_{coagulation} + S^q_{sticking}$$

where $\rho_l$ is the charge density in section l.

## IV. RESULTS

In this article we are interested in analyzing the interplay between the main mechanisms that control the particle dynamics, i.e. nucleation, cluster deposition (sticking) on particle surface and particle coagulation. We especially focus on investigating how the sputtering yield ($Y$) that controls the carbon quantity released in the discharge affect the interplay between these processes and the resulting time-evolution of the aerosol dynamics.

The base-values for the model parameters used in the simulation are a cluster sticking coefficient on the particle surface of 1, a size of 15 carbon for the largest molecular edifice, a cold electron temperature of 0.1 eV [12], a gas temperature of 300 K and an activation energy for cluster agglomeration processes of 0.225 eV. Simulations are realized for two values of the sputtering yield 1% and 2 %. The particle size distribution is investigated over a size range between 1 and 100 nm using 100 sections in the discretization of the aerosol dynamic equations.

Fig. 3 shows space-size distribution of the particle cloud in the inter-electrode region for a discharge duration of 250 s. The particle cloud is located near the field reversal position located at 2 cm from the cathode. Particle nucleation creates small particles during all the discharge duration. These particles grow by coagulation and cluster sticking on their surface. After a first growth phase, depletion takes place for the size range between 2 and 10 nm. This depletion is due to the coagulation which occurs between the small and large particles. This coagulation is possible due to the small averaged-charge carried by particles with sizes in the range 1-10 nm (around -1 to -2 elementary charge) and for which the charge fluctuation results in a 25% of neutral and positively charged particles as shown in Fig 4. Note that the scale for the particle size is logarithmic in order to point out that the particle in the 1-10 n range play a key role in the coagulation. A linear representation would show that particle charge varies linearly with the size.

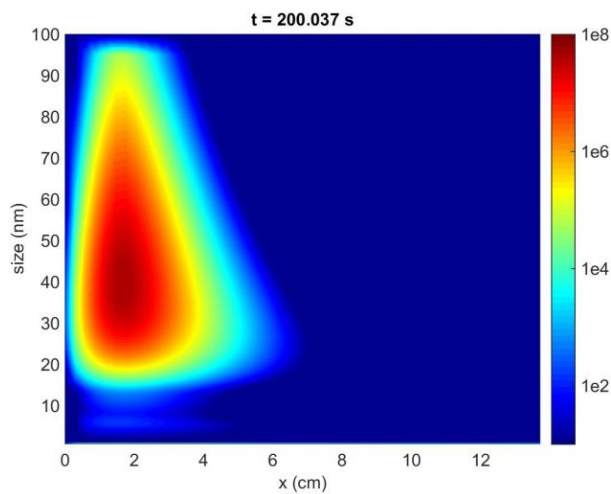

**Fig. 3 Particle size distribution at 200s for a sputtering yield of 2%.**

We also observe that the particle trapping gets more efficient as the particle size increase. Particle cloud is observed between the sheath edge at 0.3 cm from the cathode toward 6 cm for the particle smaller than 30 nm but is reduced to 4 cm large for the 100 nm particles due to their higher charge, around -12 averaged charges.

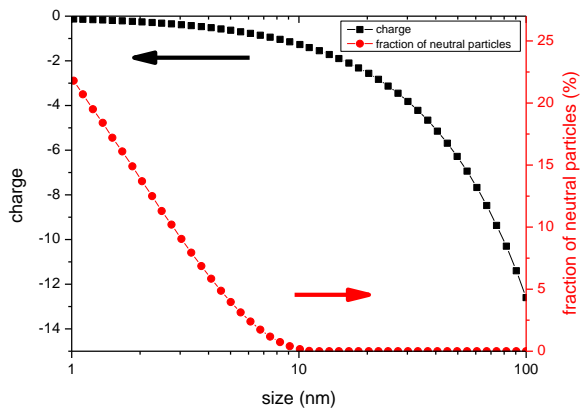

**Fig. 4:  Particle averaged charge and fraction of particle carrying an instantaneous neutral or positive charge; note that the scale for particle size is logarithmic.**

Fig.5 shows the time evolution of the particle total density at the position of the maximum density, i.e., the field reversal position. As may be expected, particles are produced faster when the carbon quantity sputtered at the cathode is higher. Basically, the increase of the sputtering yield from 1 to 2% leads to a two-fold reduction of the time needed to achieve a steady state density. The steady state particle density at the field reversal position also increases by almost a factor two from $2.2 \times 10^8$ to $4.2 \times 10^8$ cm$^{-3}$.

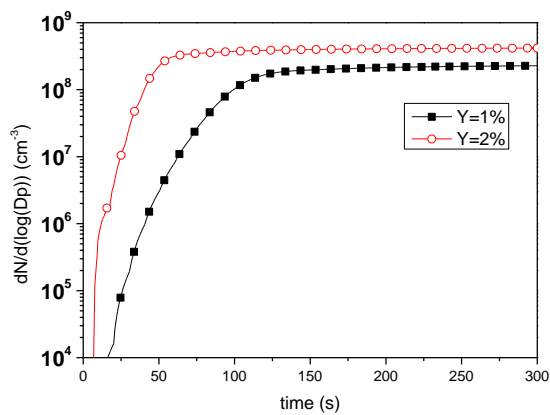

**Fig. 5 : particle density for sputtering yield of 1% (square) and 2% ( circle)**

Fig. 6 shows the time evolution of the particle mean-diameter at the field reversal position for 1% and 2% sputtering yield. Particle mean-diameter keeps increasing during the whole discharge duration. It is characterized by a first phase where the mean-diameter increases exponentially with time, and a second phase where the diameter growth kinetics is linear. The exponential phase is much faster and the linear growth phase is reached earlier for high sputtering yield, *i.e.* 2%. The linear growth phase shows the same dynamics, *i.e.*, same slope, for the two values of the sputtering yield. As a result, the difference between the mean-diameter obtained for the two sputtering yield is in the range 10-15 nm at the end of the exponential growth phase and keeps constant during the linear growth phase.

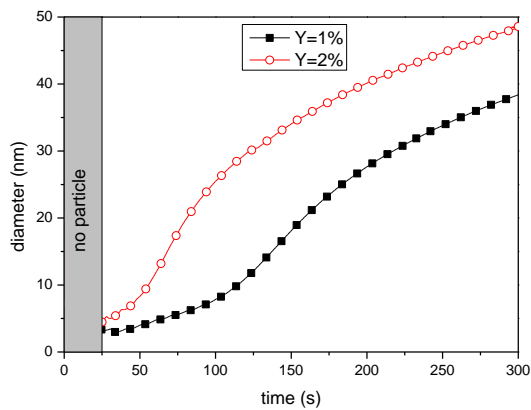

**Fig. 6 : averaged particle diameter for sputtering yield of 1% (square) and 2% ( circle)**

In order to investigate the mechanism that is responsible of this behavior. We have plotted in Fig 7 the time evolution of the particle nucleation rate at the field reversal position, for 1% and 2% sputtering yields.

Increasing the sputtering yield leads to the production of more primary C, $C_2$ and $C_3$ clusters at the cathode. As a consequence, the cluster aggregation phase, which is a second order kinetics, is enhanced and a larger number of maximum size clusters, and therefore nucleus, is produced. For 2% sputtering yield, maximum nucleation rate is reached 50 s earlier and is three-fold larger than in the case of 1% sputtering yield.

Fig 8 shows the particle size distribution calculated at the field reversal position and 300 s discharge duration for 1% and 2% sputtering yields. We also report in the same figure the particle size distribution calculated for 2% sputtering yield at 200s discharge duration. This shows the same mean-diameter as the particle size distribution obtained at 300 s for 1% sputtering yield. The first remarkable finding is that all these distributions show similar bimodal behavior with two pronounced maxima at small size, around the nucleus diameter, and large size, around the mean diameter, and a very pronounced depletion at intermediate size, i.e., 2-10 nm. Due to the faster cluster growth phase, particle distribution obtained at 2% sputtering yield shows larger populations at larger sizes. Nevertheless, particle distributions with the same mean diameter show similar shapes, *i.e.*, relative variations, no matter the value of the sputtering yield, *i.e.*, particle distributions obtained at 300s for $Y=1$% is similar to the one obtained at 200 s for $Y=2$%. Of course, the absolute particle density is two times higher for 2% sputtering yield.

We have previously shown that models that do not take into account the details of the size distribution tends to predict that particle growth is mainly dominated by cluster sticking on particle surface [3]. We will first analyze this growth term when a detailed size distributions, such as those shown in Fig. 8, is taken into account. Fig 9 shows the rates of molecular deposition in term of the mass deposited on particle for both sputtering yield values at the field reversal position. In each case we observe a first increase of the mass deposited on the particle surface that ends with a pronounced maxima at approximately 50 s for $Y=2$% and 100 s for $Y=1$%, which are consistent with the maxima observed for the nucleation rate, fig. 7. This first phase is followed by a decrease of the rate of carbon sticking on the particle surface. In this second phase, both particle density and nucleation rate are important, so that the cluster density decreases. For the largest sputtering yield, $Y=2$%, the mass rate of carbon deposited on the particle is even smaller than in the case of $Y=1$% at the end of the discharge duration. However, if we compare the total mass of cluster sticking on the particle, we can notice that this mass increase by almost a factor of two when the sputtering yield is larger, from $5 \cdot 10^{-14}$ g for $Y=1$% to $8 \cdot 10^{-14}$ g for $Y=2$%.

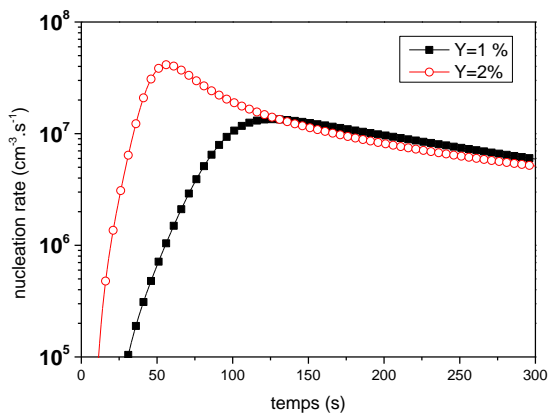

**Fig. 7 Particle nucleation rate for sputtering yield of 1% (square) and 2% ( circle)**

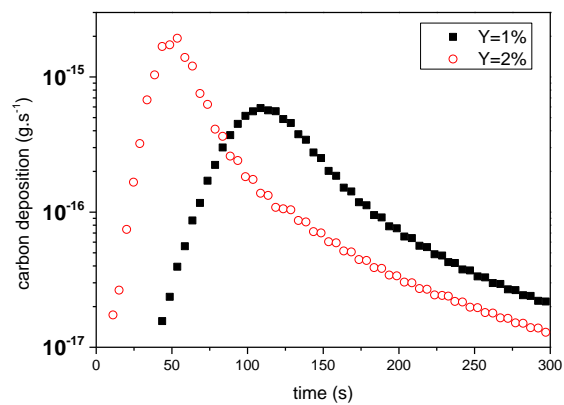

**Fig. 9 : net mass rate of carbon deposition by sticking at the position of the field reversal for sputtering yield of 1 % and 2%.**

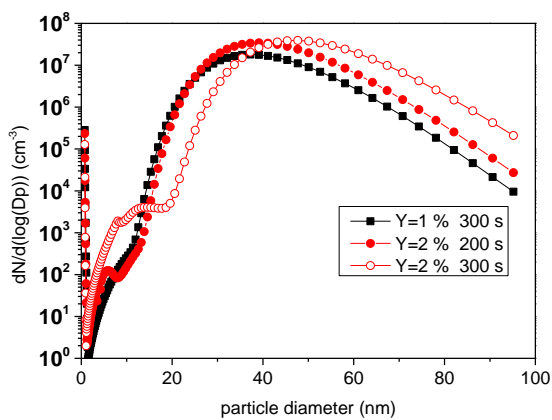

**Fig. 8 Particle density at the field reversal position at 300 s for sputtering yield of 1% at 300 s (square) and 2 % (circle) at 200s and 300 s.**

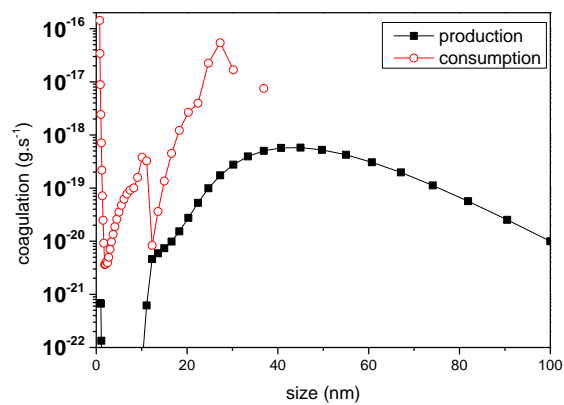

**Fig. 10 Rates of mass production (square) and consumption (circle) through coagulation process at 300 s at the position of the field reversal for sputtering yield of 1 %.**

Fig 10 shows the coagulation rates in terms of the particle mass consumed and produced at the field reversal position for $Y$=1% at 300 s . We can see that particle consumption by coagulation presents two maxima. This means that there are two particle sizes that are strongly involved in the coagulation processes. The first one is located at the smallest size, near the nucleus size. The second maximum is located at the maximum of the size distribution, (cf. fig. 8). So coagulation occurs between small particles that shows the highest probability for neutral or positive charge (fig. 4), with the predominant larger particles that are negatively charged.

Fig. 11 presents the net mass consumption of particles by coagulation for $Y$=1% at 300 s, $Y$= 2% at 200 s and 300 s. Reaction rates for both sputtering yield are very similar in shape. If we compare the net mass consumption of particle for $Y$=1%/300 s with the one for $Y$=2%/200 s, cases that show similar particle distributions, we observe that the rate is one order of magnitude higher in the case of $Y$=2% for both the smallest and the 20-30 nm particles in the Y=2%/200 s case. Particle density is however 2 times higher in this case. The enhancement of coagulation is then due to the combination of two phenomena: (i) particle density is a little bit higher for $Y$=2% and 200 s than for $Y$=1% at 300 s and (ii) particle nucleation is almost 2 times higher in the case of the higher sputtering yield, which provides a larger number of particles that are prone to undergo a coagulation process with larger particles that belong to the core of the distribution (fig. 7).

Fig. 11 shows also that coagulation rate for the same sputtering yield, *i.e.* $Y$= 2%, decreases with time. At 300 s it is one order of magnitude smaller that at 200 s. This may be once again explained by the decrease of the nucleation rate that results from the enhanced kinetics of carbon clusters sticking on the particle surface when the particle density is larger. Small nuclei which carry very small charge play then a key role in the coagulation process due to their charge fluctuation that enables their collision with other particles with larger negative charge.

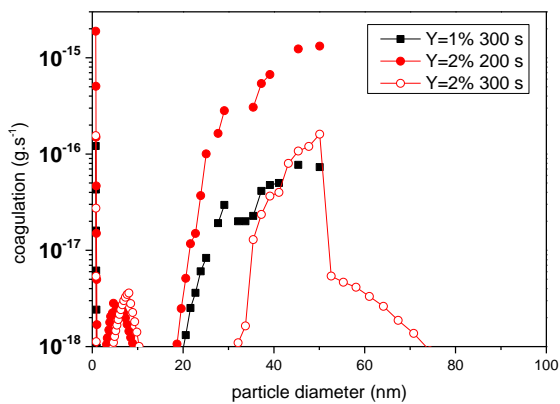

**Fig. 11 Rates of net mass consumption through coagulation for sputtering yield of 1% at 300 s (square) and 2 % (circle) at 200s and 300 s.**

## V. Concluding remarks

The present computational study makes possible obtaining a fairly detailed information on particle formation due to graphite cathode sputtering in an argon DC discharges. The modeling procedure described in this work enables us to estimate and investigate the time-evolution of some of the particle cloud characteristics, e., g. total density, mean-diameter, average- charge. We could also reach a quite detailed description of the particle size distribution that showed a bimodal behavior. We showed how the interplay between the different phenomena involved in the aerosol dynamics in dusty plasma is affected by the sputtering yield. We especially showed that when we increase the number of primary carbon cluster produced t the cathode, we enhance the particle nucleation. Then, once particle density reach $10^8$ cm$^{-3}$, the clusters produced are more consumed by the sticking on the particle surface, which limits the molecular growth kinetics and therefore the nucleation rate.

We have also showed that each single particle experiences a much more limited growth through cluster sticking on surface for large sputtering yield. On the opposite, the net mass production through coagulation is much larger in this case, and since coagulation process is much more efficient than cluster sticking in terms of particle growth, we finally observe an enhanced particle growth dynamics for the larger sputtering yields.

This shows how the strong coupling between the different phenomena involved in the aerosol dynamics, i.e. , nucleation, cluster sticking and coagulation, may result in unexpected trends when acting on one of the discharge parameters.

**Armelle Michau** received the  PhD degree from Paris 13 University, France, in 2012. She is with LSPM CNRS laboratory, Université Paris 13, France. Her main research interests include plasma modeling.

**Khaled Hassouni**  received the PhD degree from Paris 6 University, France, in 1992.
He is with LSPM CNRS laboratory, Université Paris 13, France. His main research interests include plasma modeling.

**Cécile Arnas**  received the PhD degree from Toulouse 3 University, France, in 1987.
She is with LPIIM CNRS laboratory, Université Aix Marseille, France. Her main research interests include plasma physics and nanoparticle formation.

**Guillaume Lombardi** received the PhD degree from Paris 11 University, France, in 2003. His main research interests are focused on plasma processes diagnostics.